\documentclass{emulateapj}
\usepackage{amsmath}
\usepackage{amssymb}
\usepackage{natbib}
\usepackage{xspace}
\usepackage{graphicx}
\usepackage{subfigure}
\usepackage{rotating} 
\usepackage{float}
\usepackage{comment}
\usepackage{hyperref}

\newcommand{\nustar}{\textsl{NuSTAR}\xspace}
\newcommand{\bat}{\textsl{Swift}/BAT\xspace}

\shortauthors{Kamraj et al.}

\begin{document}
	
\title{Coronal Properties of \bat-selected Seyfert 1 AGN observed with \nustar}

\author{N. Kamraj\altaffilmark{1}, F. A.~Harrison\altaffilmark{1}, M.~Balokovi\'c\altaffilmark{2}, A. Lohfink\altaffilmark{3}, M. Brightman\altaffilmark{1}}

\altaffiltext{1}{Cahill Center for Astronomy and Astrophysics, California Institute of Technology, Pasadena, CA 91125, USA} 

\altaffiltext{2}{Harvard-Smithsonian Center for Astrophysics, 60 Garden Street, Cambridge, MA 02138, USA}

\altaffiltext{3}{Department of Physics, Montana State University, 211 Montana Hall, Bozeman, MT 59717, USA}

\email{Contact: nkamraj@caltech.edu}

\begin{abstract}

\noindent The \nustar observatory, with its high sensitivity in hard X-rays, has enabled detailed broadband modeling of the X-ray spectra of Active Galactic Nuclei (AGN), thereby allowing constraints to be placed on the high-energy cutoff of the X-ray coronal continuum. We investigate the spectral properties of a sample of 46 \nustar-observed Seyfert 1 AGN selected from the \bat 70-month hard X-ray survey. Our measurements of the high-energy cutoff of the continuum from modeling the \nustar X-ray spectra are used to map out the temperature -- compactness ($\theta-l$) plane for AGN coronae. We find that most of the coronae lie clustered near the boundary for runaway pair production, suggesting that annihilation and pair production act to regulate the temperature of the corona. We discuss the implications of coronae whose high-energy cutoff may indicate a low coronal temperature on the heating and thermalization mechanisms in the corona.   
	
\end{abstract}


\section{Introduction}\label{sec:intro}

The continuum X-ray emission from Active Galactic Nuclei (AGN) is believed to originate in a hot, compact corona located above the accretion disk \citep[e.g.,][]{corona-haardt}. Compton upscattering of UV and optical photons from the inner accretion disk by coronal electrons produces a power-law-like X-ray continuum, with a cutoff at energies determined by the electron temperature $T_{e}$ \citep[e.g.,][]{rybicki-lightman,Zdziarski-2000}. The shape of the coronal continuum is sensitive to properties such as the seed photon field, electron temperature, optical depth, and observer viewing angle.  The observed rapid variability of the 2--10 keV emission in many AGN, combined with X-ray spectral timing and reverberation mapping, strongly indicate that the corona is physically compact, of the order 3--10 gravitational radii \citep{fabian-2009, kara-reverb,Emmanoulopoulos-2014,  fabian-2015}. The gravitational radius is defined to be $GM_{\rm BH}/c^{2}$, where  $M_{\rm BH}$ is the supermassive black hole mass. Such radiatively compact sources can exchange significant energy between particles and photons, with the compactness characterised by the dimensionless parameter \emph{l} \citep{guilbert-1983}, defined as:

\begin{center}
	\vspace{-15pt}
	\begin{equation}  
	l = 4\pi \frac{m_{p}}{m_{e}} \frac{R_{g}}{R} \frac{L}{L_{E}}
	\end{equation}
\end{center}
where $m_{p}$ and $m_{e}$ are the proton and electron mass respectively, $R_{g}$ is the gravitational radius, $R$ the source radius, $L$ the source luminosity, and $L_{E}$ the Eddington luminosity. The electron temperature $T_{e}$ can also be characterised by the dimensionless parameter $\theta = k_{B}T_{e}/m_{e}c^{2}$, where $k_{B}$ is the Boltzmann constant. For sufficiently energetic photons, photon-photon collisions can lead to electron-positron pair production in the corona \citep{svensson-1982, guilbert-1983, Zdziarski-1985}. At high coronal temperatures, when the Wien tail of the power-law spectrum extends above $~ 2m_{e}c^{2}$, pair production can quickly become a runaway process, exceeding annihilation \citep{svensson-1984}. This will limit any further rise in temperature, thus acting as an \emph{l}-dependent thermostat \citep{svensson-1984, Zdziarski-1985, stern-1995}. 

The \nustar observatory \citep{nustar-harrison}, being the first focusing hard X-ray telescope in orbit, has enabled detailed, high signal-to-noise spectra to be obtained in the 3--79 keV band for many local AGN. \nustar spectral modeling can thus place constraints on the spectral photon index and high-energy cutoff of the coronal X-ray continuum, enabling robust estimates of $l$ and $\theta$. One of the primary goals of the \nustar mission is to perform an extragalactic survey of the hard X-ray sky, in order to characterise the AGN population. We define hard X-rays as photons with energies $>$ 10 keV. As part of its Extragalactic Legacy Surveys program\footnote{https://www.nustar.caltech.edu/page/legacy\_surveys}, the \nustar observatory has performed snapshot $\sim$ 20 ks observations of local AGN detected in the all-sky survey with the Burst Alert Telescope (BAT) instrument onboard the \emph{Neil Gehrels Swift Observatory} \citep{swift-mission,swift-survey}. Though previous work has provided broad constraints on the high-energy cutoff for samples of bright AGN, tight constraints for particular AGN only became available recently thanks to \nustar \citep[e.g.,][]{ballantyne-2014,brenneman-2014,keck-2014,fabian-2015,mislav-2015}. The 100-fold increase in sensitivity of the \nustar telescope compared to the \bat instrument enables robust spectral modeling with a minimal \nustar exposure of $\sim$ 20 ks. With even longer exposure \nustar observations, it is possible to obtain tight limits on X-ray spectral parameters and perform reverberation mapping measurements of coronal size.  

In this paper, we study a sample of 46 \bat selected Seyfert 1 (Sy1) AGN observed with \nustar, in order to map out the location of these sources on the temperature -- compactness ($\theta-l$) diagram for AGN coronae. We do not include \emph{Swift}/XRT data in our spectral modeling as the limited data quality of available simultaneous \emph{Swift}/XRT data introduces difficulties in obtaining constraints on parameters such as the cutoff energy. The complexity of features in soft X-ray spectra would ideally require high signal-to-noise ratio, simultaneous spectra from soft X-ray telescopes with larger collecting area to model robustly, which are currently unavailable for the targets in our sample. In section~\ref{sec:data}, we discuss the sample used in this study, the data reduction, and analysis procedures adopted. Observational details of our AGN sample are presented in Table 2 of the appendix. In section~\ref{sec:results}, we present our results and discuss the heating and cooling mechanisms operating in the corona. We discuss future, deeper \nustar observations of AGN in our sample with potential cutoffs in the \nustar band in section~\ref{sec:futureobs}, and present a summary in section~\ref{sec:summary}. In this work, all uncertainties were calculated at the 90\% confidence level and standard values of the cosmological parameters ($h_{0} = 0.7$, $\Omega_{\Lambda} = 0.7$,  $\Omega_{m} = 0.3$) were used to calculate distances.

\section{Sample, Data Reduction, and Analysis}\label{sec:data}

\subsection{Sample of Seyfert 1 AGN}

We selected our sample from AGN identified in the \bat 70--month hard X-ray catalogue \citep{swift-mission,swift-survey}. From the full catalogue, we selected \nustar-observed AGN with known redshifts and classified as Sy1 from optical hydrogen emission line measurements, or from available data from the NASA/IPAC Extragalactic Database (NED). The full list of AGN included in our study, along with their \nustar observation details, may be found in Table 2 of the appendix. Figure 1 shows the location of our sources on the redshift-luminosity plane, with the luminosity values determined from the \bat fluxes in the 14--195 keV range. We confirmed sources at high redshift to not be beamed AGN or blazar candidates from observations of their optical spectra and cross-matching with the Roma Blazar Catalog \citep{roma-bzcat}. We found two sources which were misclassified from NED and were removed from our sample. We excluded 9 sources from our original sample due to lack of constraints on the high-energy cutoff from spectral fitting. Our final sample consists of 46 Sy1 AGN at $0.003 < z < 0.2$. 

In Figure 2, we present the distributions of \bat fluxes, luminosities and redshifts for both our sample and the Sy1 classified sources from the \bat 70-month catalog. We find that our sample is statistically representative of the Sy1 population from the \bat 70-month catalog, with the mean and median values overlapping between our sample and the parent \bat sample. We further applied a two-sample Kolmogorov-Smirnov test and found the K-S test statistic to be 0.1 or lower, and the p-value above 60 \% for all three distributions, thus confirming that the distributions are consistent between our sample and the larger \bat sample of Sy1s.    

\begin{figure}[t]
	\hspace{-8pt}
	\includegraphics[width=0.5\textwidth]{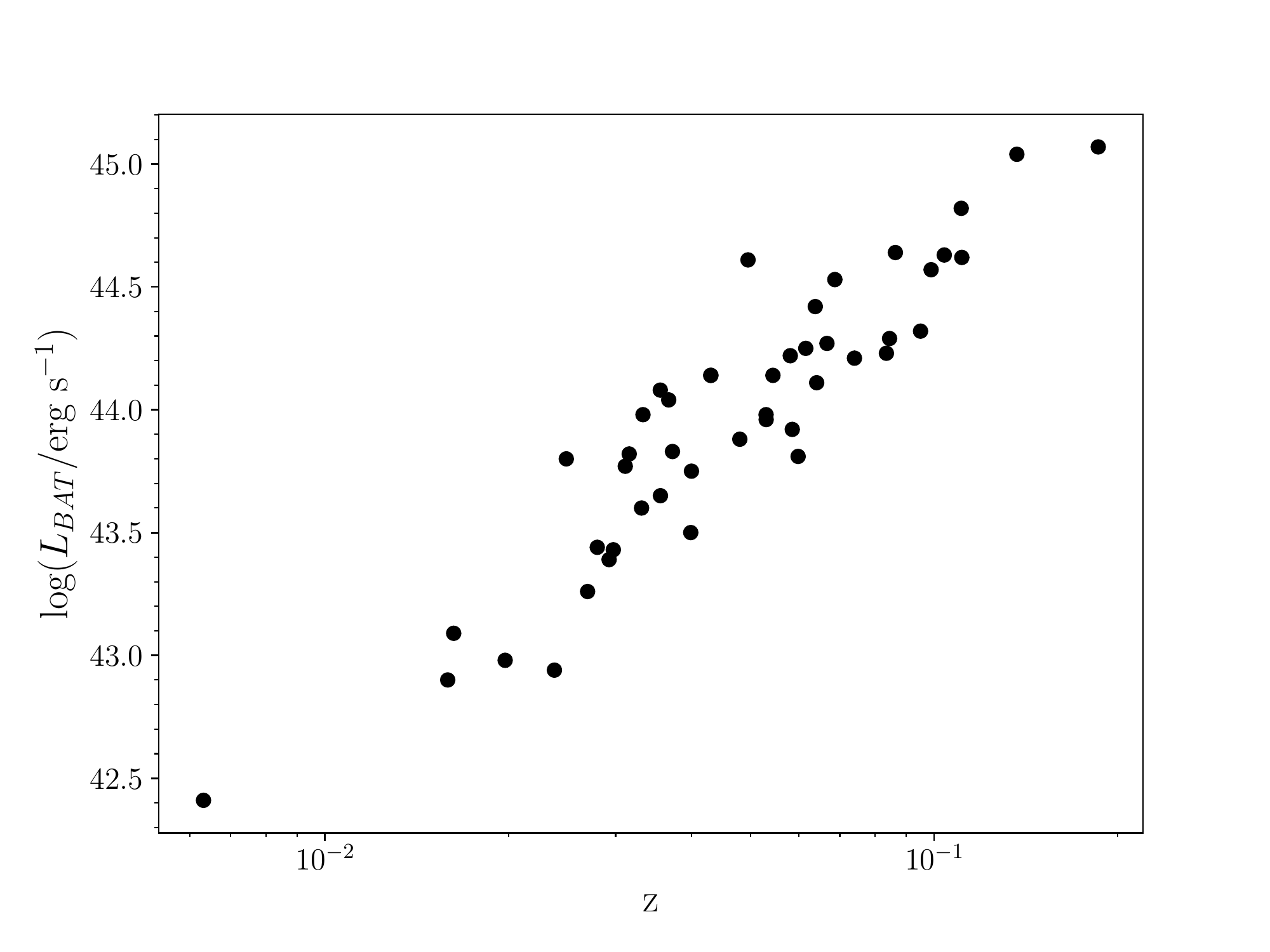}
	\label{fig1}
	\vspace{-8pt}
	\caption{Redshift-luminosity distribution for \nustar-observed Sy1 AGN selected from the \emph{Swift}/BAT 70-month hard X-ray catalogue.}
\end{figure}        

\begin{figure}[t]
	\centering   
	\begin{subfigure}{}
		\hspace{-20pt}
		\includegraphics[width=0.5\textwidth]{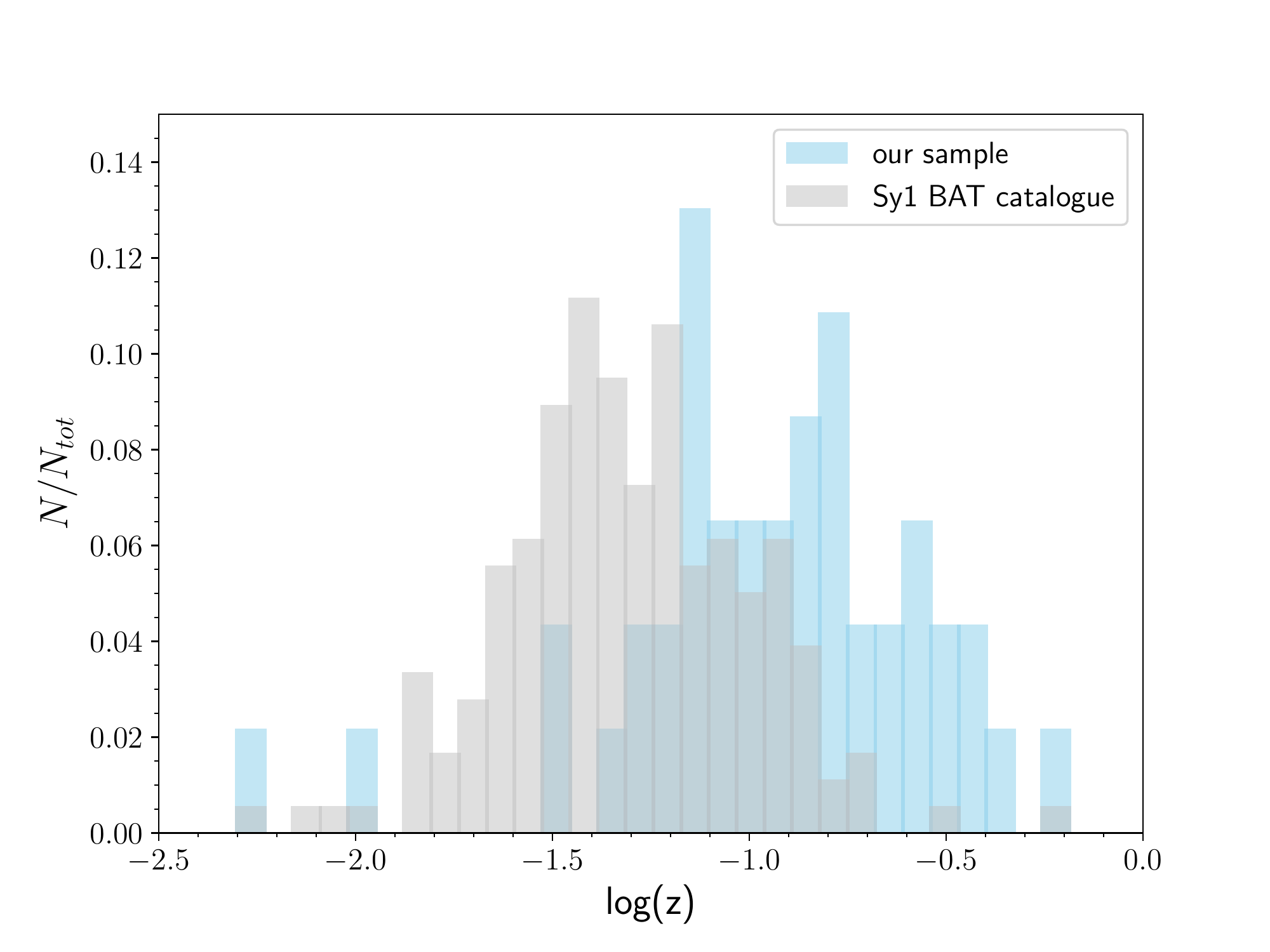}
	\end{subfigure}%
   \begin{subfigure}{}
   	   \hspace{-20pt}
       \includegraphics[width=0.5\textwidth]{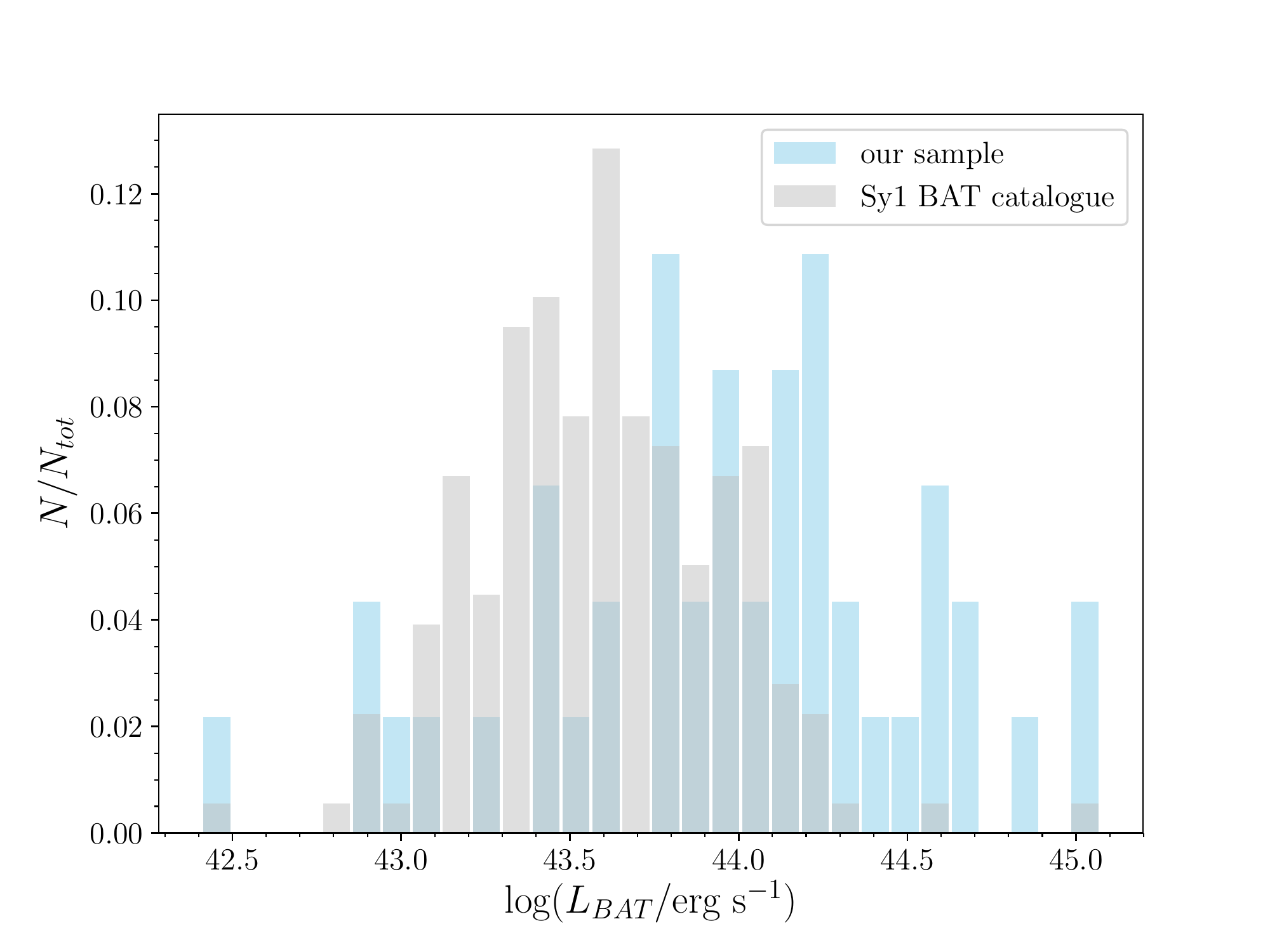}
   \end{subfigure}%
   \begin{subfigure}{}
   	 \hspace{-20pt}
     \includegraphics[width=0.5\textwidth]{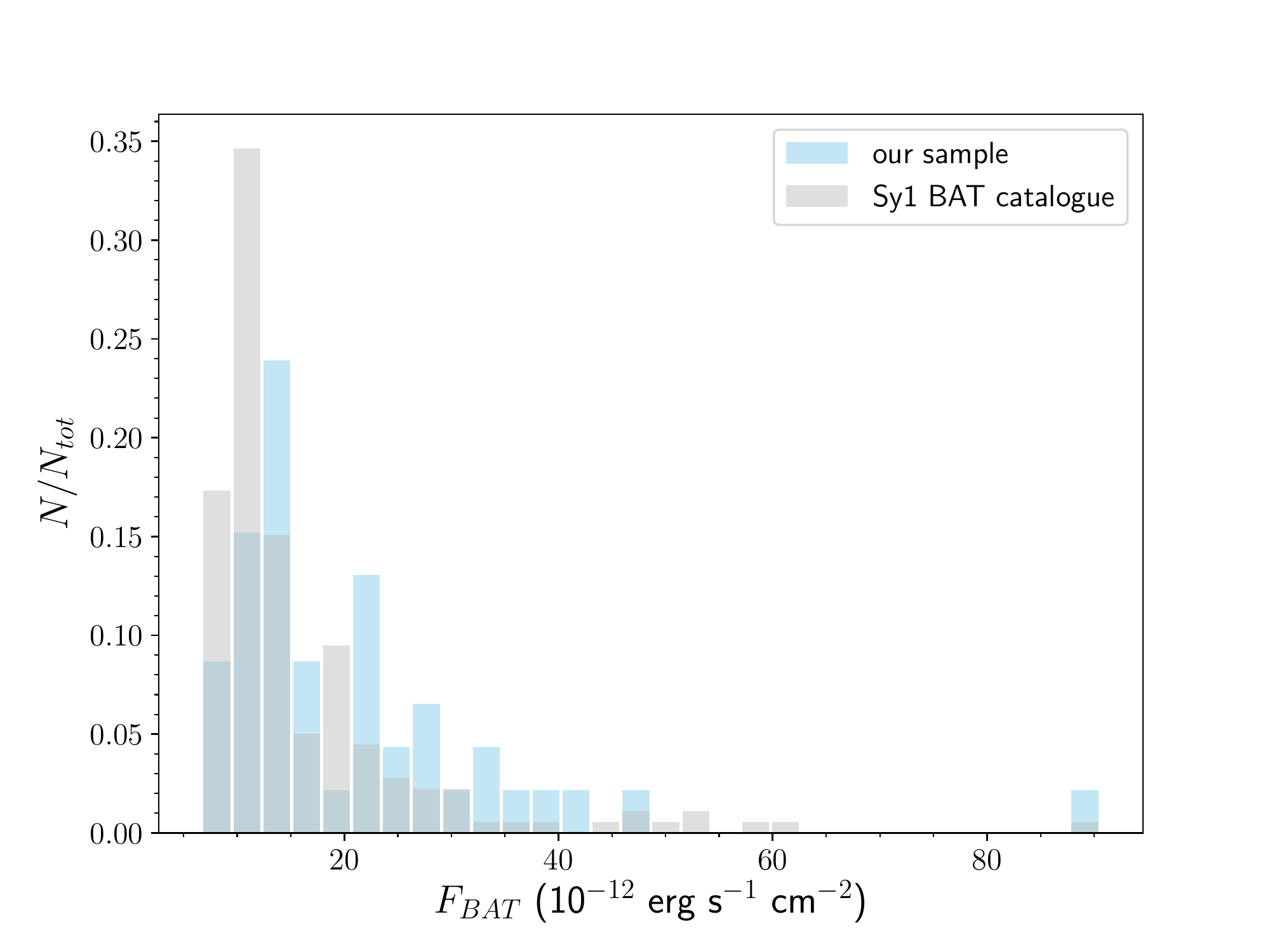}
   \end{subfigure}%
	\caption{Distributions of redshifts, \bat 70-month X-ray catalogue luminosities and fluxes for both our sample and the Sy1 classified sources from the \bat catalogue. For clarity, sources with $L_{BAT} < 10^{40}$ erg s$^{-1}$ and $F_{BAT} > 1\times10^{-10}$ erg s$^{-1}$ cm$^{-2}$ were omitted from the plots.}
	\label{fig:figure}	
\end{figure}

\subsection{NuSTAR Observations and Data Reduction}

 Roughly once per week since its start of science operations in
 2013, the \nustar satellite has been obtaining $\sim$ 20 ks observations
 in the 3--79~keV band of AGN selected from the \bat 70-month hard X-ray catalog \citep{swift-survey}. We performed reduction of raw event data from both \nustar modules, FPMA and FPMB \citep{nustar-harrison}, using the \nustar Data Analysis Software (NuSTARDAS, version 1.2.1), distributed by the NASA High Energy Astrophysics Archive Research Center (HEASARC) within the HEASOFT package, version 6.16. We took instrumental responses from the \nustar calibration database (CALDB), version 20160502. Raw event data were cleaned and filtered for South Atlantic Anomaly (SAA) passages using the \texttt{nupipeline} module. We extracted source and background energy spectra from the calibrated and cleaned event files using the \texttt{nuproducts} module. Detailed information on these data reduction procedures can be found in the \nustar Data Analysis Software Guide \citep{nustardas}. An extraction radius of 30\arcsec\ was used for both the source and background regions. We extracted the background spectrum from source-free regions of the image, and away from the outer edges of the field of view, which have systematically higher background. The spectral files were rebinned using the HEASOFT task \texttt{grppha} to give a minimum of 20 photon counts per bin. For multiple observations of the same source, we coadded spectra using the HEASOFT task \texttt{addspec}.

\subsection{Spectral Modeling}

We performed spectral modeling of the \nustar data in the 3--79 keV band for each source in our sample using XSPEC v12.8.2 \citep{xspec}. We used $\chi^{2}$ statistics for all model fitting and error estimation. We adopted cross sections from \citet{vern} and solar abundances from \citet{wilm}. In all our modeling we include a cross-correlation constant between FPMA and FPMB to account for slight differences in calibration \citep{Madsen:2015aa}

We fit each spectrum with an absorbed power-law model with a high-energy cutoff, $E_{\rm cut}$. The slope of the power-law continuum is characterized by the photon index, $\Gamma$. It is assumed that the instrinsic continuum intensity is proportional to $E^{-\Gamma}$exp($-E/E_{\rm cut}$). In XSPEC notation, the model used is \texttt{TBabs $\times$ zwabs $\times$ cutoffpl}, where the component \texttt{TBabs} models Galactic absorption, which is fixed to a typical Galactic column density of $7.6\times10^{21}$ cm$^{-2}$ \citep{galactic-nh}. We found that freezing the Galactic column density did not have any significant effect on the fit results, as spectral modeling over the hard X-ray band is relatively insensitive to this parameter. The redshifted component \texttt{zwabs} accounts for absorption by the host galaxy. 

Where an Fe K$\alpha$ emission line feature was observed in the spectra at 6.4 keV, we added an additive \texttt{zgauss} Gaussian line component to the absorbed power-law model. We note that two objects out of our sample required fitting with an Fe K$\alpha$ line: Mrk 595 and RBS 1037. In addition, we test for the presence of spectral features due to reprocessing by adding a \texttt{pexrav} component \citep{pexrav}. We fixed elemental abundances to solar and kept the inclination angle fixed at the default value of {60\degr}. We found that the reduced $\chi^{2}$ values and best-fit parameters from modeling with \texttt{pexrav} were similar to those from fitting an absorbed cutoff power-law for the majority of the sources in our sample, indicating that the addition of a reflection component does not significantly modify fit results and thus is not required by the data. Futhermore, we found that the null hypothesis probability exceeds 50 \% for many of our sources when fitting with an absorbed cutoff power-law; we found the mean null hypothesis probability of our sample to be 43 \%. We note that we chose a reflection model for one source (2MASX J19301380$+$3410495) due to best-fit parameters such as the photon index being more physically reasonable compared to the absorbed cutoff power-law model. We also note that the reduced $\chi^{2}$ for Mrk 9 is relatively high due to increased scatter in the data near $\sim$ 10 keV and 30 keV, which do not correspond to any known physical features. We summarize some of the key best-fit spectral parameters for our sample in Table 3 of the appendix. We did not find any sources in our sample with significant line of sight absorption ($>5\times10^{23}$ cm$^{-2}$), with most sources having hydrogen column densities constrained to be $<10^{22}$ cm$^{-2}$. Figure 3 presents an example \nustar spectrum for a potential low-cutoff candidate in our sample, 2MASX J19301380$+$3410495, for which we measured $E_{\rm cut}$ to be 23$^{+29}_{-9}$ keV.

\begin{figure}[t]
	\hspace{-15pt}
	\includegraphics[width=0.52\textwidth]{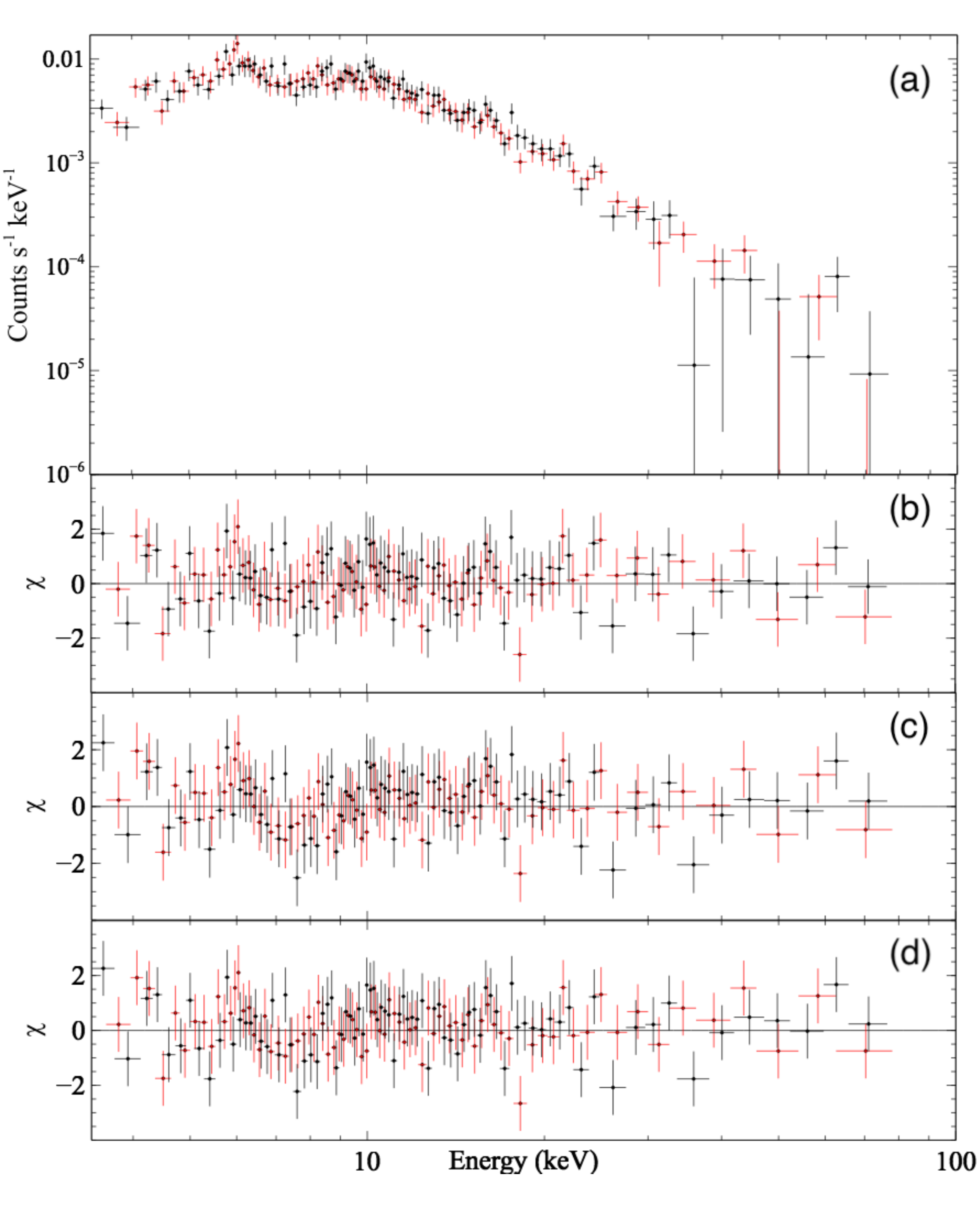}
	\caption{\nustar hard X-ray spectrum of a candidate low cutoff AGN 2MASX J19301380$+$3410495 (a), alongside fit residuals for (b) an absorbed power-law model ($\chi^{2}$/dof = 155.1/161) (c) absorbed power-law model with a high-energy cutoff ($\chi^{2}$/dof = 144.4/160) and (d) absorbed cutoff power-law with reflection modeled via \texttt{pexrav} ($\chi^{2}$/dof = 138/160). Black points correspond to FPMA data while points in red correspond to FPMB.}
	\vspace{5pt}
\end{figure}

\section{Results and Discussion}\label{sec:results}

In this section, we present limits on the high-energy cutoff, $E_{\rm cut}$, found from spectral modeling of our sample. We then present the location of our sources on the $\theta-l$ plane for AGN coronae, and discuss the implications of sources with low values of $E_{\rm cut}$ on the heating and cooling mechanisms operating in the corona. 

\subsection{Cutoff Constraints}

The distribution of lower limits on the high energy cutoff for our sample is presented in Figure 4. The histogram shows a number of AGN with lower limits on $E_{\rm cut}$ below 100 keV. Typical values of $E_{\rm cut}$ for AGN generally range from $\sim$ 100 keV to 300 keV \citep{dadina-07, malizia-2014,ricci-2017a}; we note \citet{x-ray-background} comment that the mean value of $E_{\rm cut}$ for AGN must not exceed several hundred keV, in order to avoid overproducing the cosmic X-ray background above 100 keV.  

\begin{figure}[t]
	\includegraphics[width=0.5\textwidth]{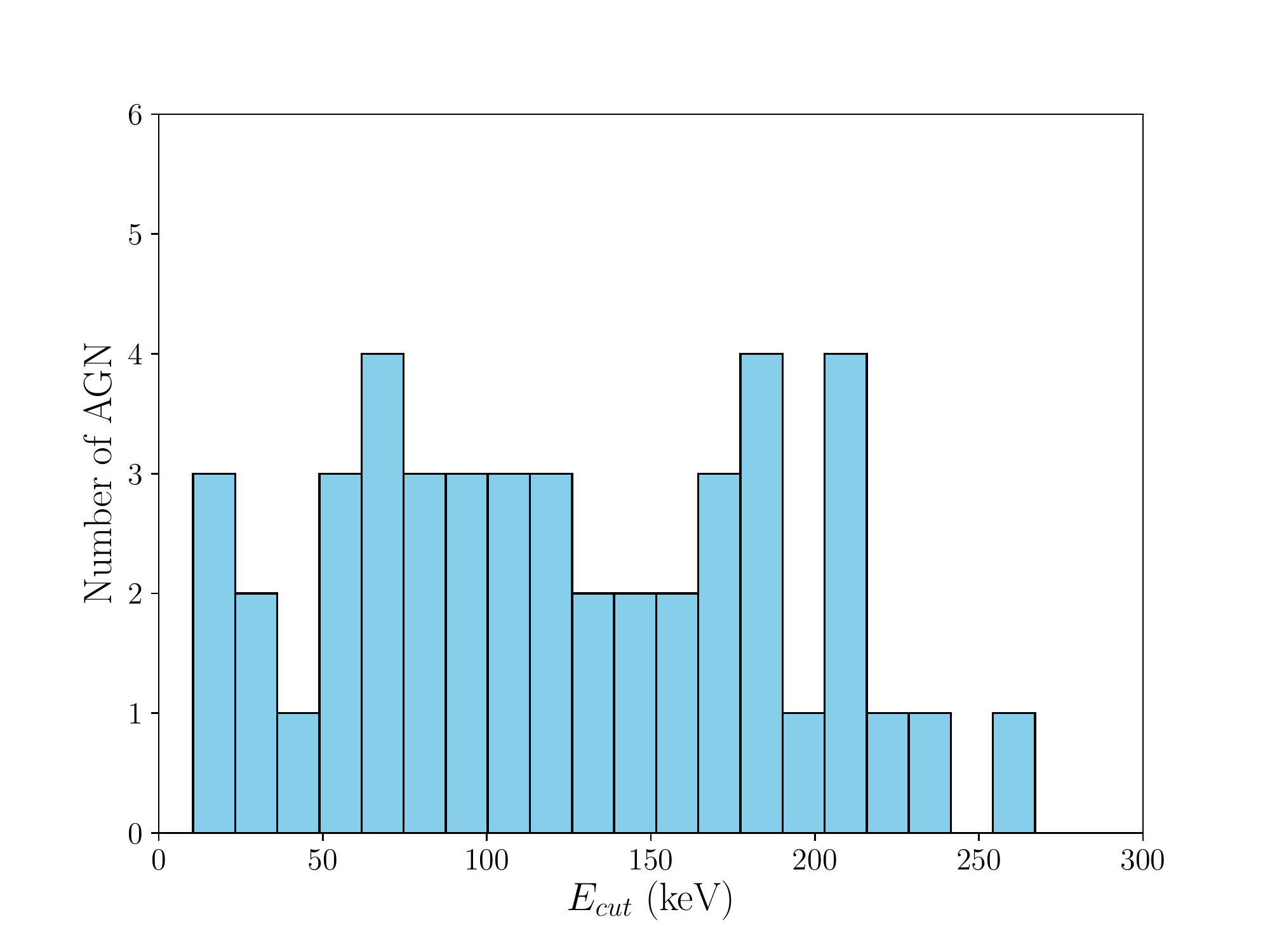}
	\label{fig2}
	\caption{Distribution of lower limits on $E_{\rm cut}$ for our Sy1 AGN sample from modeling \nustar data.}
	\vspace{3pt}
\end{figure}    

Despite the fact that the quality of \nustar data in the hard X-ray band exceeds any previous observations of our targets, the cutoff power-law model does still display a degree of degeneracy in the derived photon index ($\Gamma$) and $E_{\rm cut}$. In order to verify that our constraints on $E_{\rm cut}$ are physically reasonable, in Figure 5 we compare our derived $\Gamma$ and $E_{\rm cut}$ values to curves of constant optical depth in the $E_{\rm cut}$--$\Gamma$ parameter space. The purple line in Figure 5 corresponds to theoretical constraints from \citet{petrucci-2001} for an optical depth $\tau = 6$. We use the relationship derived for a slab geometry of the corona by Petrucci et al. (2001) to calculate the optical depth as a function of $\Gamma$ and $E_{\rm cut}$:

\begin{center}
	\begin{equation}  
	\Gamma = \sqrt{\frac{9}{4}+\frac{511~\mbox{keV}}{\tau kT_{e}(1+\tau/3)}} - \frac{1}{2}.
	\end{equation}
\end{center}

AGN coronae are typically thought to be optically thin ($\tau<1$) \citep{Zdziarski-1985,stern-1995}, though some have been constrained to $\tau \sim 3$ based on high-quality \nustar data \citep[e.g.,][]{mislav-2015,grs-low-ecut,kara-2017}. Combinations of $\Gamma$ and $E_{\rm cut}$ that correspond to $\tau>6$ can be considered to result from a degeneracy between model parameters and therefore are unphysical. With this particular assumption we suspect that for 3 targets our results may be unrealistic; if, for example, $\tau<10$ is chosen, then no targets fall in this category. However, sources lying near or below the line with $\tau = 6$ were not removed from our sample, as the limited \nustar data quality with a short, 20 ks exposure does not rule out physically reasonable values of the photon index.

\begin{figure}[t]
	\hspace{-10pt}
	\includegraphics[width=0.52\textwidth]{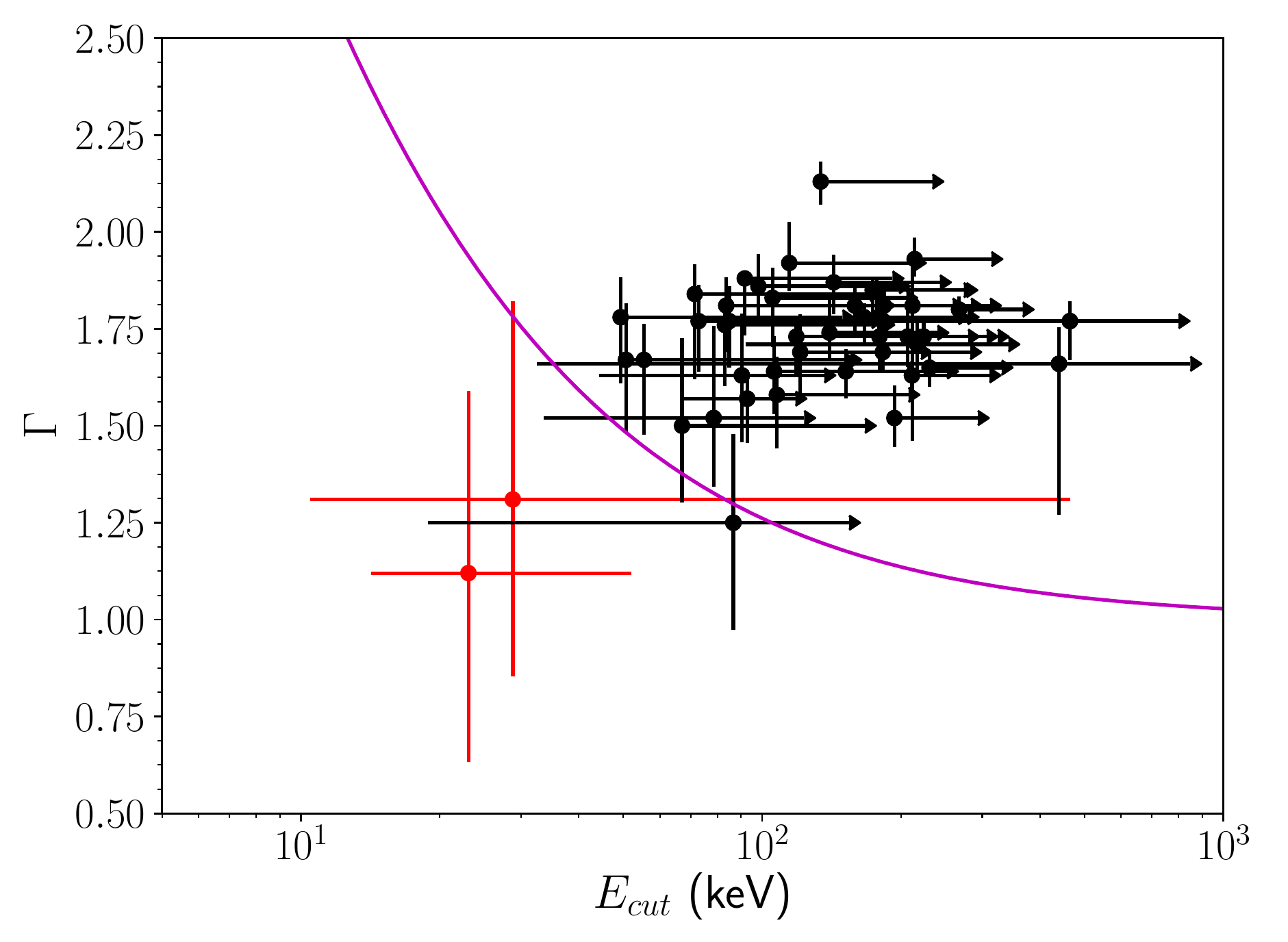}
	\label{fig3}
	\caption{$E_{\rm cut}$ vs photon index $\Gamma$ for our sample. Points in red denote candidate sources with low coronal high-energy cutoffs for which both upper and lower limits on $E_{\rm cut}$ were measured. The purple line corresponds to theoretical constraints from \citet{petrucci-2001} for $\tau = 6$.}
	\vspace{6pt}
\end{figure} 

We investigate the presence of model degeneracies in the sources with the lowest measured $E_{\rm cut}$ constraints (2MASX J19301380$+$3410495 \& 1RXS J034704.9-302409) by exploring the $E_{\rm cut}$--$\Gamma$ parameter space in XSPEC. Figure 6 shows the contour plots of the photon index against the high-energy cutoff for these sources. Whilst there is some degree of degeneracy between these two parameters, the value of $E_{\rm cut}$ is constrained to low values over the range of physically reasonable photon index values at the 68 \% confidence level.


 \begin{figure}[t]
 	\begin{subfigure}{}
 	\hspace{-30pt} 
 		\includegraphics[width=0.5\textwidth]{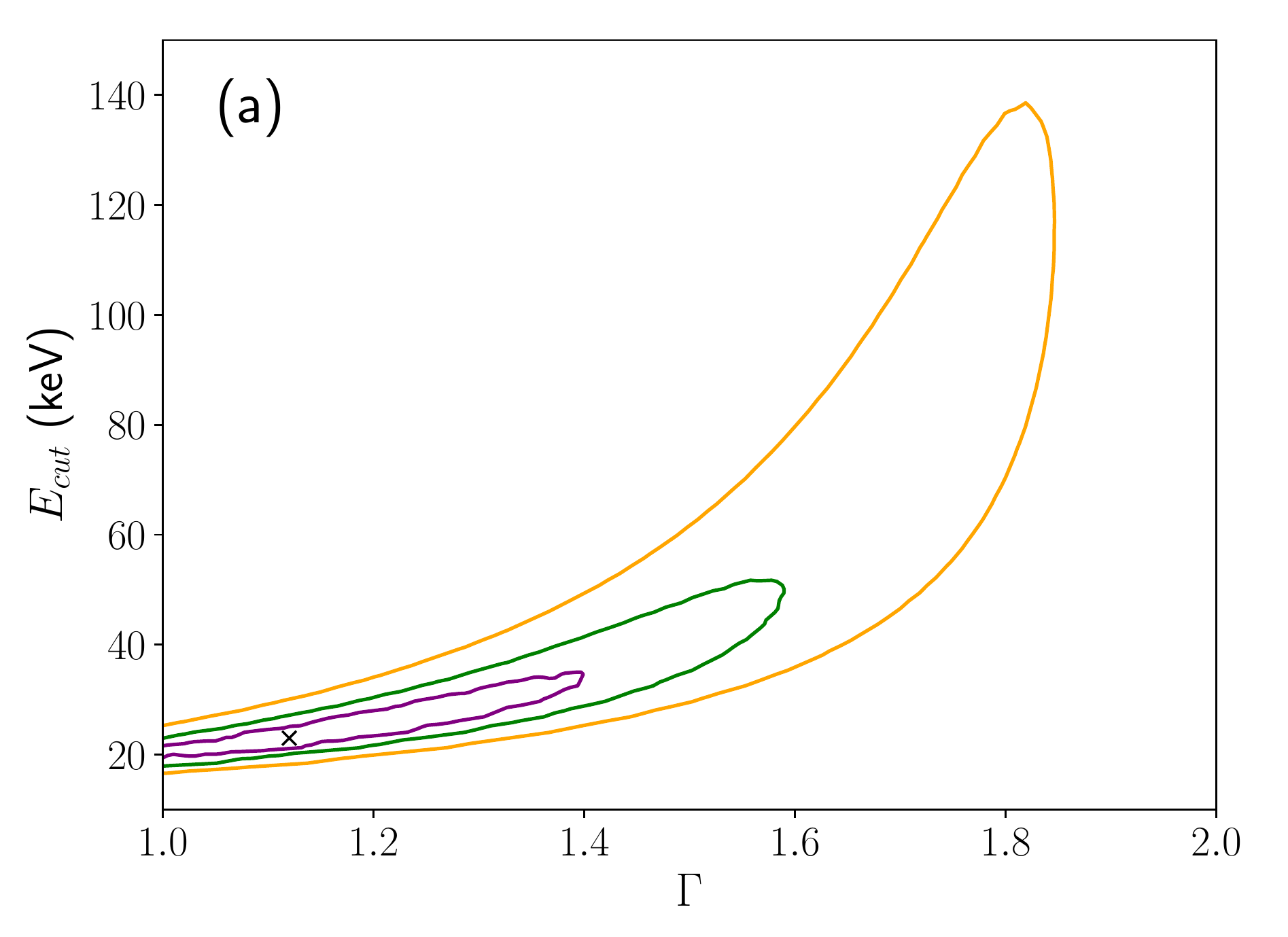}
 	\end{subfigure}%
 	\begin{subfigure}{}
 	\hspace{-20pt} 
 		\includegraphics[width=0.5\textwidth]{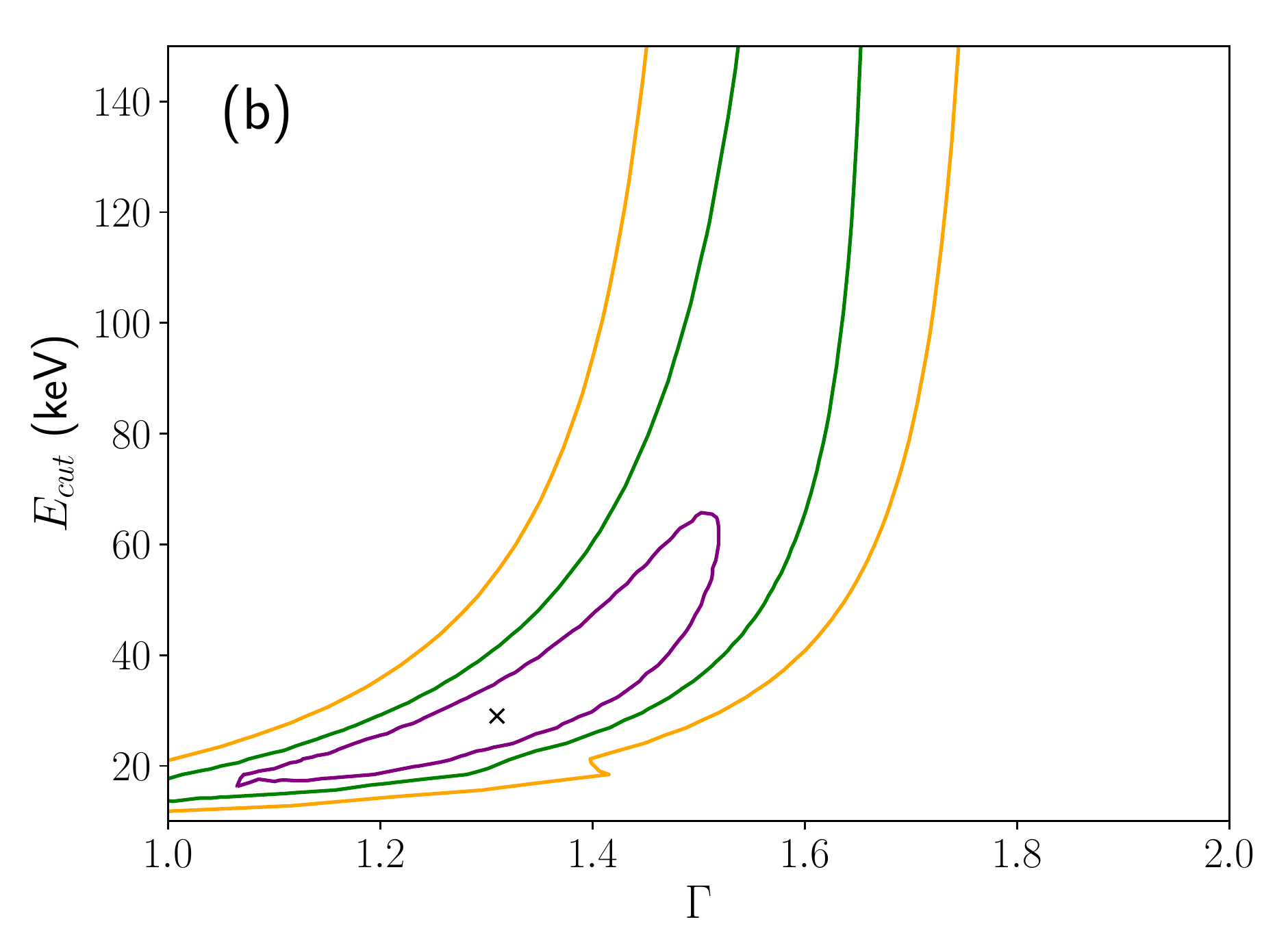}
 	\end{subfigure}%
 	\caption{$E_{\rm cut}$--$\Gamma$ contour plots for \nustar observations of the candidate low cutoff Sy1s (a) 2MASX J19301380$+$3410495 (b) 1RXS J034704.9-302409. The solid purple, green and yellow contours correspond to the 68, 90 and 99 \% confidence levels, respectively. The black cross represents the best fit values of the parameters from applying the relevant model given in Table 3.}
 	\vspace{5pt}
 	\label{fig:figure}	
 \end{figure} 

\subsection{The $\theta-l$ Plane}

In constructing the observational $\theta-l$ plane, we convert from $E_{\rm cut}$ to the coronal temperature using $k_{B}T_{e} = E_{\rm cut}/2$ \citep{petrucci-2001}. In calculating $l$, we assume a conservative value of 10$R_{g}$ for the coronal radius $R$, as adopted in \citet{fabian-2015}, as the majority of the sources in our sample lack the required X-ray reflection modeling or reverberation measurements to place constraints on coronal size. We estimated the source luminosity $L$, from the flux in the 0.1--200 keV band, which was extrapolated from the applied spectral model. We convert the unabsorbed 0.1--200 keV flux obtained from spectral modeling to luminosity using luminosity distance values from NED. Black hole mass estimates, where available, were taken from \citet{koss-2017}. The values of $M_{\rm BH}$ used in \citet{koss-2017} were obtained from a combination of broad Balmer emission line measurements, direct techniques such as X-ray reverberation mapping, and the $M_{\rm BH}-\sigma_{\ast}$ relation of \citet{m-sig-relation}. 
We have black hole mass measurements obtained from the literature for 34 of the 46 sources in our sample. For sources with no published black
hole mass, we use the median black hole mass of the Sy 1--1.5 AGN in the BAT AGN Spectroscopic Survey (BASS) \citep{koss-2017}, log($M_{\rm BH}/M_{\odot}$) $=$ 7.97$\pm$0.52.

We note that the precise location of AGN on the $\theta-l$ plane is dependent on general relativistic effects, such as gravitational redshift and light bending. Processes such as light bending introduce inclination-dependent corrections to $l$. These corrections depend on the geometry of the corona, which is currently highly uncertain. Therefore, due to the large uncertainties associated with model-based relativistic corrections, we do not include general relativistic effects here. 

Figure 7 presents the location of our sources on the $\theta-l$ plane, in addition to theoretical pair lines for different coronal geometries. Runaway pair production occurs to the right of the pair lines, as described in the introduction. Modeling the corona as an isolated electron cloud, \citet{svensson-1984} estimated the pair production line to have the analytical form

\begin{center}
	\vspace{-20pt}
	\begin{equation}  
	l \sim 10\theta^{5/2} e^{1/\theta}.
	\end{equation}
\end{center}

\citet{stern-1995} also computed the pair balance line for a slab and hemispherical corona respectively, located above a reflecting accretion disk. The solid black and purple lines in Figure 7 correspond to these geometries. 

\begin{figure}[t]
	\centering
	\hspace{-30pt}   
	\includegraphics[width=0.52\textwidth]{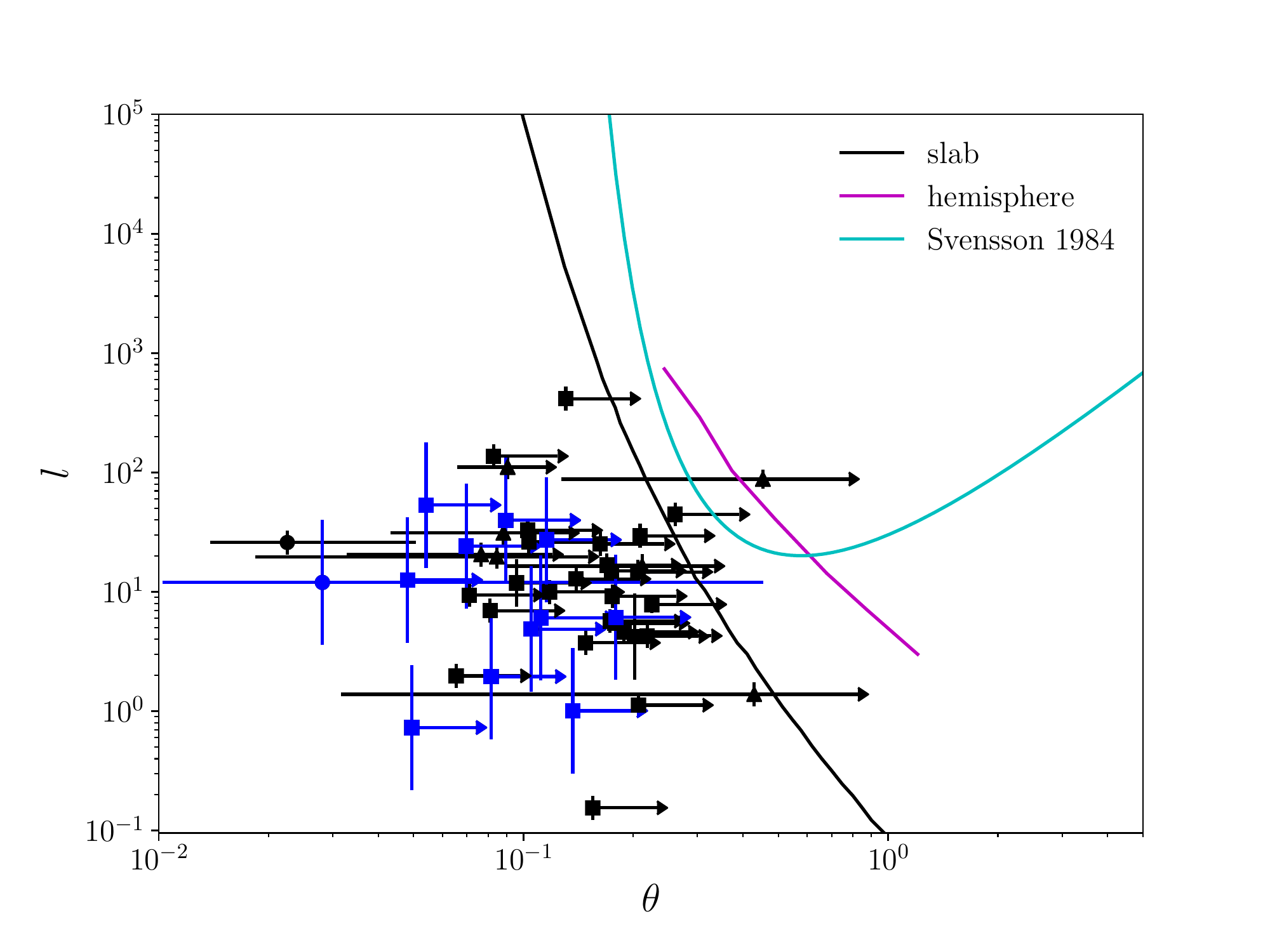}
	\caption{The $\theta-l$ plane for \nustar-observed Sy1 AGN. Solid lines correspond to pair lines for different coronal geometries. Circled points are candidate low cutoff sources for which both upper and lower limits on $E_{\rm cut}$ were measured. Triangles denote sources with a best-fit value and lower limit on $E_{\rm cut}$. Squares denote sources with only lower limits on $E_{\rm cut}$. Blue points indicate sources for which the black hole mass was taken to be the median black hole mass of the type 1 AGN in the BAT AGN Spectroscopic Survey (BASS) \citep{koss-2017}.}
	\vspace{5pt}
\end{figure}

Our results show that most of the AGN coronae in our sample are clustered near the lines for runaway pair production, similar to the results found by \citet{fabian-2015} for \nustar-observed AGN and black hole binaries (BHB). The pair lines thus appear to correspond to a physical boundary, constraining sources to that region. A few AGN are located away from the pair line boundary, hinting at low coronal temperatures. Note that we have assumed that the corona is homogeneous and at a single temperature, whereas in reality there may be a range of temperatures. This may result in a mean temperature at a lower value due to Compton cooling \citep{fabian-2015}.

Recent detections of low coronal cutoffs have been made within the \nustar band, such as \citet{grs-low-ecut}, \citet{kara-2017} and \citet{yanjun-2017}. For example, \citet{kara-2017} measured $T_{e} = 15\pm2$ keV for the narrow-line Sy1 Ark 564, making it one of the lowest temperature coronae observed by \nustar to date. Multiple explanations have been proposed for the origin of low temperature coronae. In the case of an AGN accreting close to the Eddington limit, the stronger radiation field may enhance Compton cooling in comparison with sub-Eddington Seyferts \citep{kara-2017}. For sources accreting well below the Eddington limit, the relatively low coronal temperatures may be attributed to highly effective cooling in some AGN due to, e.g., high spin and the resulting higher seed photon temperature. Low temperatures may also arise from particularly weak coronal heating mechanisms, or more effective cooling due to multiple scatterings in a corona with high optical depth \citep[e.g.,][]{grs-low-ecut}. Naively, when the optical depth in the corona exceeds unity, multiple inverse Compton scatterings transfer a proportionally higher fraction of the stored thermal energy to coronal luminosity. However, coronae are complex systems, and many coupled physical processes determine the electron temperature. 

Another possibility is that the corona consists of a hybridized plasma, containing both thermal and non-thermal particles \citep[e.g.,][]{Zdziarski-1993,corona-heating-paper,fabian-2017}. In such a system, the corona is highly magnetized and compact, and thus heating and cooling are so intense that electrons do not have time to thermalize before they are cooled by inverse Compton scattering. The presence of only a small fraction of non-thermal electrons with energies above 1 MeV can result in runaway pair production. The cooled electron-positron pairs may redistribute their available energy, thereby reducing the mean energy per particle and decreasing the coronal temperature. Such cooling would produce a hard non-thermal tail and an annihilation feature at 511 keV. Hard X-ray data of very high quality are necessary to distinguish between a hybrid, pair-dominated plasma and cooler, fully thermal plasma incapable of pair production.

\section{Future Observations}\label{sec:futureobs}

\begin{table*}
	\centering	
	\caption{Mean values of the high-energy cutoff, its lower and upper limits from simulated \nustar spectra, for \bat-selected Sy1 AGN}
	\renewcommand{\arraystretch}{1.5}
	
	\begin{tabular*}{\textwidth}{@{\extracolsep{\fill} }l c c c c c}
		\noalign{\smallskip} \hline \hline \noalign{\smallskip}	
		Name & Exposure Time & $E_{\rm cut}$ & $E_{\rm cut}$ lower limit & $E_{\rm cut}$ upper limit \\
		& (ks) & (keV) & (keV) & (keV) \\ \hline	
		1RXS J034704.9-302409 & 50 & 92.7 & 17.1 & 118.8 \\
		& 100 & 54.6 & 19.2 & 111.8 \\
		2MASX J19301380+3410495 & 50 & 21.4 & 15.7 & 33.0 \\
		& 100 & 20.7 & 16.7 & 27.0 \\
		Mrk 1393 & 50 & 189.0 & 43.5 & 191.7 \\
		& 100 & 155.0 & 45.0 & 199.7 \\
		SDSS J104326d47+110524.2 & 50 & 170.4 & 46.3 & 203.2 \\
		& 100 & 134.1 & 52.1 & 206.5 \\
		UGC 06728 & 50 & 162.6 & 59.4 & 234.8 \\
		& 100 & 127.5 & 67.6 & 222.8 \\	
		\\ \hline
	\end{tabular*}
	\vspace{25pt}	
\end{table*}

The $E_{\rm cut}$ constraints presented here are based on
snapshot $\sim 20$~ks \nustar observations of a sample of
bright Sy1 galaxies, and identified several sources which potentially
have high-energy cutoffs within the \nustar band (i.e., 3--79
keV). Future work will involve performing longer exposure \nustar observations of AGN from our sample that display hints of a low coronal cutoff, which will aid in removing model degeneracies and more tightly constrain $E_{\rm cut}$, in order to determine the coronal temperature. In choosing AGN from our sample for longer exposure \nustar observations, we performed 5000 simulations of the spectra of candidate low $E_{\rm cut}$ AGN from our sample in XSPEC, for exposure times of 50 ks and 100 ks. From the simulated spectra, we plotted distributions of the best-fit value of $E_{\rm cut}$ found from applying an absorbed cutoff power-law model, in addition to lower limits and upper limits on $E_{\rm cut}$. The plots in Figure 8 show distributions of values of $E_{\rm cut}$ for one such candidate low cutoff source, 2MASX J19301380$+$3410495. Table 1 summarizes the mean values of $E_{\rm cut}$ and its lower and upper limits obtained from our simulations for some candidate low cutoff AGN in our sample.

\begin{figure}[t]
	\centering  
	\begin{subfigure}{}
		\centering
		\includegraphics[width=0.5\textwidth]{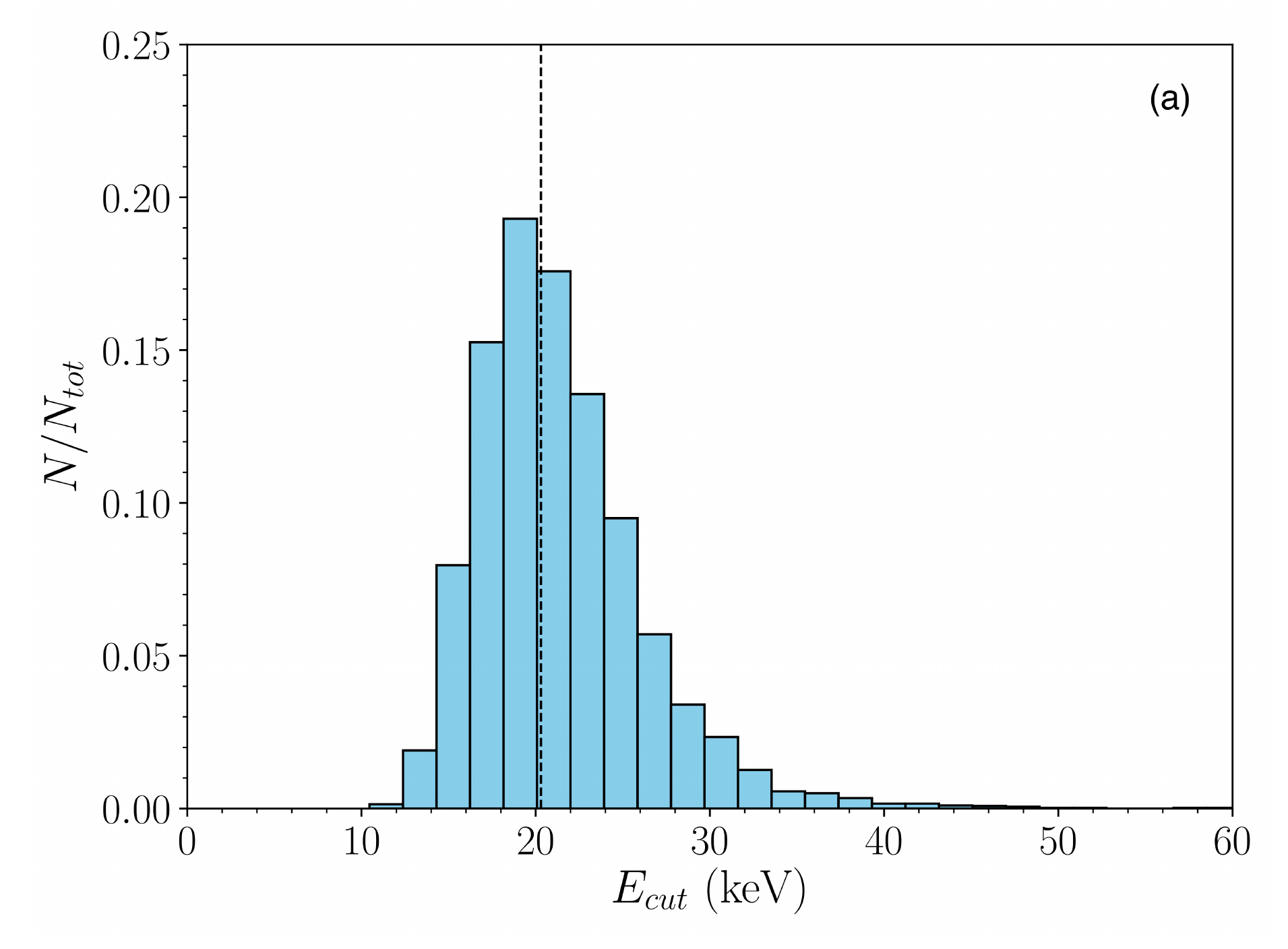}
	\end{subfigure}%
	\begin{subfigure}{}
		\centering
		\includegraphics[width=0.5\textwidth]{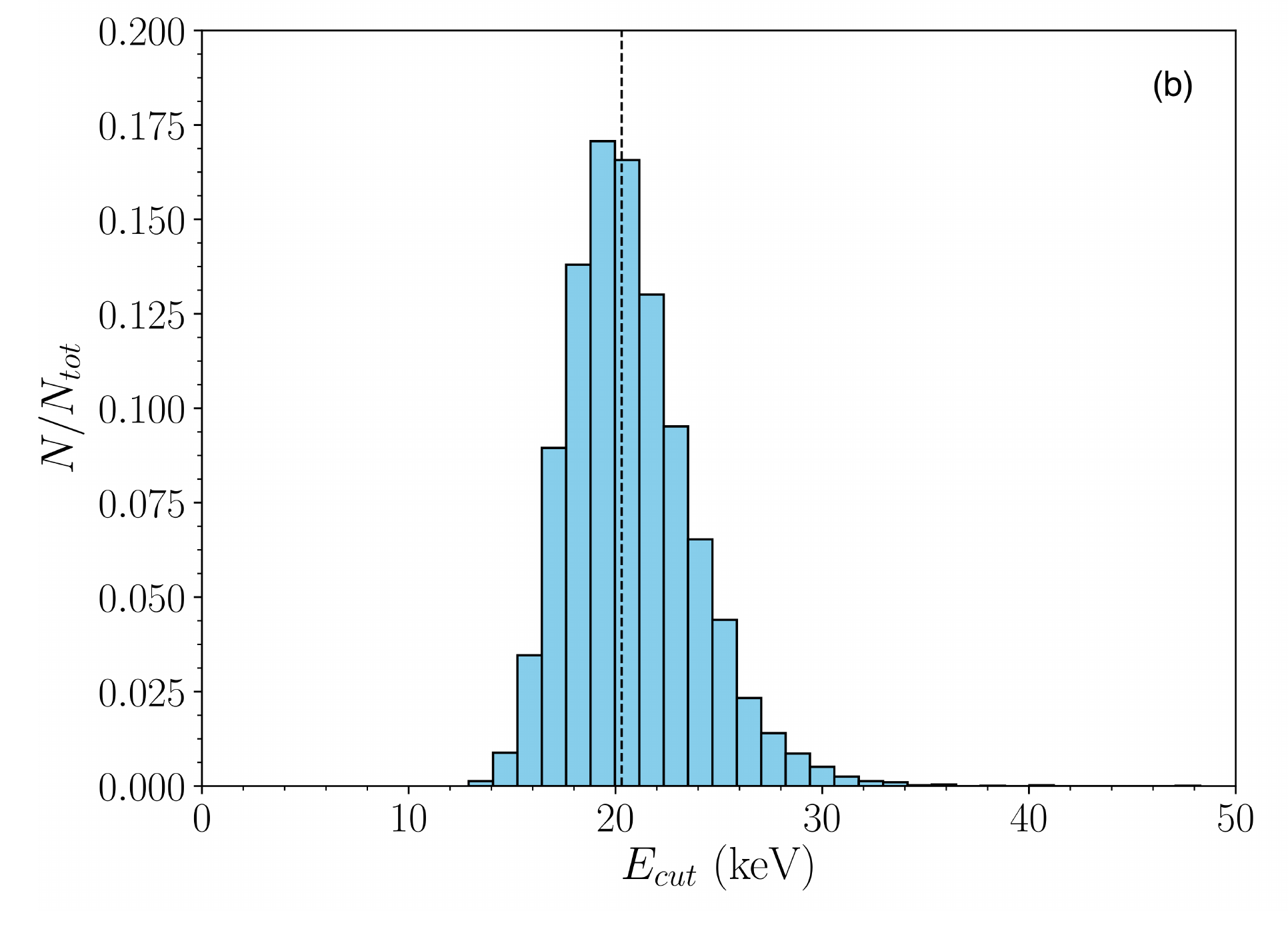}
	\end{subfigure}%
	\caption{Distributions of $E_{\rm cut}$ values for 5000 simulations of the \nustar spectrum of a candidate low cutoff AGN, 2MASX J19301380$+$3410495. Dotted lines denote the input $E_{\rm cut}$ value assumed for simulated spectra. Simulations were performed for exposure times of (a) 50 ks and (b) 100 ks.}
	\label{fig:figure}	
\end{figure} 

The simulation results show that the distributions peak at low values of $E_{\rm cut}$, and at similar values for both a 50 ks and 100 ks exposure. We conclude from our results that a 50 ks exposure should be sufficient to constrain the high-energy cutoff in our sample of candidate low cutoff AGN.

\section{Summary}\label{sec:summary}

In this work, we have investigated the coronal properties of a sample of \emph{Swift}/BAT selected Seyfert 1 AGN that have been observed with \nustar. We individually modeled the \nustar spectra of all sources in our sample and searched for sources with hints of low coronal cutoffs, observable within the \nustar band. We mapped out the location of the sources in our sample on the compactness-temperature diagram for AGN coronae, and found that the majority of sources lie near the boundary for runaway pair production. The pair production line corresponds to a physical boundary, constraining AGN to that region. A few AGN located away from the pair lines may possess low coronal temperatures; deeper 50 ks \nustar observations will be performed of these sources to constrain the coronal temperature and optical depth. The detection of low coronal cutoffs may be explained via scenarios such as a strong radiation field, large optical depth, or a hybrid pair-dominated plasma. Further computations with hybrid plasma models, taking into account general relativistic effects, will help elucidate in more detail the heating and cooling mechanisms operating in the corona.    

\vspace{15pt}
 
We would like to thank the referee for their helpful comments, which helped improve the final version of the manuscript. We have made use of data from the \nustar mission, a project led by the California Institute of Technology, managed by the Jet Propulsion Laboratory, and funded by the National Aeronautics and Space Administration. We thank the \nustar Operations, Software and Calibration teams for support with the execution and analysis of these observations. This research has made use of the \nustar Data Analysis Software (NuSTARDAS) jointly developed by the ASI Science Data Center (ASDC, Italy) and the California Institute of Technology (USA). M. Balokovi\'{c} acknowledges support from NASA Headquarters under the NASA Earth and Space Science Fellowship Program (grant NNX14AQ07H) and support from the Black Hole Initiative at Harvard University, through the grant from the John Templeton Foundation.

\facility{\emph{Facilities}: \nustar}

\bibliographystyle{apj}
\maketitle
\bibliography{BAT_Sy1_refs}

\appendix

\pagebreak

\begin{table*}
	\centering	
	\caption{\nustar Observation Details for \bat--selected Sy1 AGN}
	\renewcommand{\arraystretch}{1.0}
	
	\begin{tabular*}{\textwidth}{@{\extracolsep{\fill} }l c c c c c c}
		
		\noalign{\smallskip} \hline \hline \noalign{\smallskip}
		
		Name & \bat ID & Observation ID & Observation Date & Exposure Time & Total Counts \\
		& & & & (ks) & \\ \hline
		
		1RXS J034704.9-302409 & SWIFT J0347.0-3027 & 60061039002 & 2013-03-15  & 6.4 & 13.5 \\
		& & 60061039004 & 2013-03-24 & 12.7 & 20.4 \\
		& & 60061039006 & 2013-04-02 & 9.5 & 21.9 \\
		1RXS J174538.1+290823 & SWIFT J1745.4+2906 & 60160674002 & 2014-12-09  & 20.3 & 2233 \\
		1RXS J213445.2-272551 & SWIFT J2134.9$-$2729 & 60061306002 & 2013-10-22  & 19.8 & 2178 \\
	    2MASS J19334715+3254259 & SWIFT J1933.9+3258 & 60160714002 & 2016-05-31 & 12.6 & 3024 \\
	    2MASX J04372814-4711298 & SWIFT J0437.4$-$4713 & 60160197002 & 2015-12-09  & 20.0 & 1500 \\
	    2MASX J12313717-4758019 & SWIFT J1232.0$-$4802 & 60160498002 & 2016-08-21 & 19.3 & 1718 \\
	    2MASX J15144217-8123377 & SWIFT J1513.8$-$8125 & 60061263002 & 2013-08-06  & 13.3 & 1011 \\
	    2MASX J15295830-1300397 & SWIFT J1530.0$-$1300 & 60160617002 & 2017-02-14  & 24.2 & 2130 \\
	    2MASX J19301380+3410495 & SWIFT J1930.5+3414 & 60160713002 & 2016-07-19  & 20.5 & 1701 \\
	    2MASX J19380437-5109497 & SWIFT J1938.1$-$5108 & 60160716002 & 2016-07-15 & 21.8 & 2834 \\
	    2MASX J20005575-1810274 & SWIFT J2001.0$-$1811 & 60061295002 & 2016-10-25  & 21.9 & 1367 \\
	    2MASXi J1802473-145454 & SWIFT J1802.8$-$1455 & 60160680002 & 2016-05-01  & 20.0 & 6800 \\	
	    3C 227 & SWIFT J0947.7+0726 & 60061329002 & 2014-02-20 & 17.2 & 293 \\
	    & & 60061329004 & 2014-02-26 & 12.1 & 188 \\ 
	    4C +18.51 & SWIFT J1742.2+1833 & 60160672002 & 2017-03-27 & 22.5 & 1080 \\
	    ESO 438-G009 & SWIFT J1110.6$-$2832 & 60160423002 & 2015-02-01  & 21.7 & 1302 \\
	    Fairall 1146 & SWIFT J0838.4$-$3557 & 60061082002 & 2014-07-27  & 21.3 & 4473 \\
	    Fairall 1203 & SWIFT J0001.6$-$7701 & 60160002002 & 2015-04-11  & 34.1 & 1739 \\
	    {[HB89]} 0241$+$622 & SWIFT J0244.8+6227 & 60160125002 & 2016-07-31 & 23.4 & 9126 \\
	    IGR J14471-6414 & SWIFT J1446.7$-$6416& 60061257002 & 2013-05-28  & 15.0 & 975 \\
	    IGR J14552-5133 & SWIFT J1454.9$-$5133 & 60061259002 & 2013-09-19  & 21.9 & 2190 \\
	    IRAS 04392-2713 & SWIFT J0441.2$-$2704 & 60160201002 & 2015-12-20  & 19.5 & 2145 \\
	    LCRSB 232242.2-384320 & SWIFT J2325.5$-$3827 & 60160826002 & 2016-07-08  & 22.5 & 495 \\
	    Mrk 9 & SWIFT J0736.9+5846 & 60061326002 & 2013-10-29 & 23.3 & 1142 \\
	    Mrk 376 & SWIFT J0714.3+4541  & 60160288002 & 2015-04-07  & 24.2 & 1791 \\
	    Mrk 595 & SWIFT J0241.6+0711 & 60160119002 & 2017-01-18 & 21.3 & 873 \\
	    Mrk 732 & SWIFT J1113.6+0936 & 60061208002 & 2013-06-11  & 26.3 & 3419 \\
	    Mrk 739E & SWIFT J1136.0+2132 & 60260008002 & 2017-03-16  & 18.5 & 1277 \\
	    Mrk 813 & SWIFT J1427.5+1949 & 60160583002 & 2017-01-23  & 24.6 & 2952 \\
	    Mrk 817 & SWIFT J1436.4+5846 & 60160590002 & 2015-07-25  & 21.9 & 2847 \\
	    Mrk 841 & SWIFT J1504.2+1025 & 60101023002 & 2015-07-14 & 23.4 & 6084 \\
	    Mrk 1018 & SWIFT J0206.2$-$0019 & 60160087002 & 2016-02-10  & 21.6 & 583 \\
	    Mrk 1044 & SWIFT J0230.2$-$0900 & 60160109002 & 2016-02-08  & 21.7 & 2821 \\
	    Mrk 1310 & SWIFT J1201.2$-$0341 & 60160465002 & 2016-06-17  & 21.1 & 2743 \\
	    Mrk 1393 & SWIFT J1508.8$-$0013 & 60160607002 & 2016-01-19  & 22.4 & 896 \\
	    NGC 0985 & SWIFT J0234.6$-$0848 & 60061025002 & 2013-08-11  & 13.9 & 2363 \\
	    PG 0804+761 & SWIFT J0810.9+7602 & 60160322002 & 2016-04-02  & 17.3 & 1903 \\ 
	    PKS 0558-504 & SWIFT J0559.8$-$5028 & 60160254002 & 2016-11-19 & 21.0 & 2940 \\
	    RBS 0295 & SWIFT J0214.9$-$6432 & 60061021002 & 2017-01-14  & 23.3 & 1887 \\ 
	    RBS 0770 & SWIFT J0923.7+2255 & 60061092002 & 2012-12-26  & 18.9 & 6426 \\
	    RBS 1037 & SWIFT J1149.3$-$0414 & 60061215002 & 2017-02-02  & 40.7 & 2198 \\
	    RBS 1125 & SWIFT J1232.1+2009 & 60061229002 & 2016-07-28  & 20.0 & 1280 \\
	    SBS 1136+594 & SWIFT J1139.1+5913 & 60160443002 & 2014-12-26  & 23.5 & 3760 \\
	    SDSS J104326.47+110524.2 & SWIFT J1043.4+1105  & 60160406002 & 2016-06-14  & 20.1 & 137 \\
	    UGC 06728 & SWIFT J1143.7+7942 & 60160450002 & 2016-07-10 & 22.6 & 2486 \\
	    UM 614 & SWIFT J1349.7+0209 & 60160560002 & 2015-03-31 & 18.2 & 2002 \\
	    WKK 1263 & SWIFT J1241.6$-$5748 & 60160510002 & 2016-04-27  & 16.4 & 7872 \\

		\\ \hline	
		
	\end{tabular*}
	
\end{table*}

\begin{table*}
	\centering	
	\caption{Redshifts, black hole masses and best-fit spectral parameters from fitting \nustar data for our \bat-selected Sy1 AGN sample}
	\renewcommand{\arraystretch}{1.0}
	
	\begin{tabular*}{\textwidth}{@{\extracolsep{\fill} }l c c c c c c c c}
		
		\noalign{\smallskip} \hline \hline \noalign{\smallskip}
		
		Source & Redshift & log($M_{\rm BH}/M_{\bigodot}$)$^{A}$ & $\Gamma$ & $E_{\rm cut}$ & F$_{0.1-200}^{B}$ & $\chi^{2}$/dof & Model$^{C}$ \\
		& & & & (keV) & $10^{-12}$ erg cm$^{-2}$ s$^{-1}$ & \\ \hline
		
		1RXS J034704.9-302409* & 0.095 & 7.97$\pm$0.52 & 1.31$^{+0.51}_{-0.46}$ & 29$^{+437}_{-18}$ & 1.12$^{+0.38}_{-0.10}$ & 50.9/50 & 1 \\
		1RXS J174538.1+290823 & 0.111 & 8.82$\pm$0.10 & 1.76$^{+0.06}_{-0.16}$ & $\geq$  83 & 8.59$^{+0.91}_{-0.89}$ & 181.5/187 & 1 \\
		1RXS J213445.2-272551 & 0.067 & 6.99$\pm$0.10 &  1.77$^{+0.09}_{-0.12}$ & $\geq$  85 & 7.62$^{+0.73}_{-0.38}$ & 155.8/175 & 1 \\
		2MASS J19334715+3254259 & 0.057 & 7.88$\pm$0.10 &  1.78$^{+0.04}_{-0.06}$ & $\geq$  166 & 15.3$^{+0.6}_{-0.6}$ & 236.3/225 & 1 \\
		2MASX J04372814-4711298* & 0.053 & 7.97$\pm$0.52 &  1.92$^{+0.11}_{-0.07}$ & $\geq$  114 & 5.08$^{+0.56}_{-0.25}$ & 147.5/123 & 1 \\
		2MASX J12313717-4758019* & 0.028 & 7.97$\pm$0.52 &  1.81$^{+0.07}_{-0.12}$ & $\geq$  84 & 5.88$^{+0.75}_{-0.46}$ & 107.8/139 & 1\\
		2MASX J15144217-8123377 & 0.068 & 8.96$\pm$0.10 &  1.66$^{+0.09}_{-0.39}$ & $\geq$  32 & 6.67$^{+0.94}_{-1.24}$ & 95.3/86 & 1 \\
		2MASX J15295830-1300397* & 0.104 & 7.97$\pm$0.52 & 1.73$^{+0.04}_{-0.10}$ & $\geq$  119 & 5.49$^{+0.25}_{-0.25}$ & 172.4/170 & 1 \\
		2MASX J19301380+3410495 & 0.063 & 8.15$\pm$0.10 & 1.12$^{+0.47}_{-0.49}$ & 23$^{+29}_{-9}$ & 22.9$^{+5.98}_{-3.77}$ & 138/160 & 2 \\
		2MASX J19380437-5109497 & 0.040 & 7.23$\pm$0.10 & 1.83$^{+0.08}_{-0.12}$ & $\geq$  105 & 9.03$^{+0.78}_{-0.65}$ & 214.2/215 & 1 \\
		2MASX J20005575-1810274 & 0.037 & 8.07$\pm$0.36 & 1.73$^{+0.08}_{-0.08}$ & $\geq$  207 & 9.62$^{+0.79}_{-0.76}$ & 285.1/250 & 1 \\
		2MASXi J1802473-145454 & 0.003 & 7.76$\pm$0.10 & 1.81$^{+0.05}_{-0.07}$ & $\geq$  159 & 23.9$^{+1.3}_{-1.2}$ & 466.3/451 & 1 \\
		3C 227 & 0.086 & 8.61$\pm$0.10 & 1.63$^{+0.16}_{-0.17}$ & $\geq$  44 & 11.7$^{+1.1}_{-0.9}$ & 331/347 & 1 \\
		4C +18.51* & 0.186 & 7.97$\pm$0.52 & 1.67$^{+0.09}_{-0.19}$ & $\geq$  55 & 3.04$^{+0.34}_{-0.20}$ & 73.2/100 & 1 \\
		ESO 438-G009* & 0.024 & 7.97$\pm$0.52 & 1.74$^{+0.09}_{-0.07}$ & $\geq$  140 & 3.95$^{+0.23}_{-0.23}$ & 92.3/113 & 1 \\
		Fairall 1146* & 0.031 & 7.97$\pm$0.52 & 1.81$^{+0.04}_{-0.05}$ & $\geq$  184 & 14.1$^{+0.9}_{-0.5}$ & 365.9/326 & 1 \\
		Fairall 1203* & 0.058 & 7.97$\pm$0.52 & 1.58$^{+0.11}_{-0.07}$ & $\geq$  108 & 3.37$^{+0.38}_{-0.33}$ & 139.4/150 & 1 \\
		{[HB89]} 0241$+$622 & 0.044 & 8.09$\pm$0.10 & 1.63$^{+0.04}_{-0.05}$ & $\geq$  211 & 24.1$^{+1.0}_{-7.4}$ & 631.1/565 & 1 \\
		IGR J14471-6414 & 0.053 & 7.70$\pm$0.10 & 1.77$^{+0.09}_{-0.13}$ & $\geq$  73 & 4.08$^{+0.51}_{-0.28}$ & 84.3/82 & 1 \\
		IGR J14552-5133 & 0.016 & 6.86$\pm$0.10 & 1.73$^{+0.03}_{-0.09}$ & $\geq$  180 & 6.43$^{+0.35}_{-0.25}$ & 191.9/181 & 1 \\
		IRAS 04392-2713* & 0.084 & 7.97$\pm$0.52  & 1.84$^{+0.08}_{-0.22}$ & $\geq$  71 & 7.82$^{+0.71}_{-0.44}$ & 200.8/173 & 1 \\
		LCRSB 232242.2-384320* & 0.036 & 7.97$\pm$0.52 & 1.67$^{+0.14}_{-0.19}$ & $\geq$  51 & 1.44$^{+0.20}_{-0.13}$ & 50.9/46 & 1 \\
		Mrk 9 & 0.040 & 7.59$\pm$0.10 & 1.52$^{+0.08}_{-0.08}$ & $\geq$  193 & 2.83$^{+0.19}_{-0.18}$ & 155.1/100 & 1\\
		Mrk 376 & 0.056 & 8.17$\pm$0.10 & 1.64$^{+0.06}_{-0.07}$ & $\geq$  152 & 4.38$^{+0.37}_{-0.20}$ & 170.3/147 & 1 \\
		Mrk 595 & 0.027 & 8.28$\pm$0.10 & 1.50$^{+0.23}_{-0.20}$ & $\geq$  67 & 2.62$^{+0.43}_{-0.33}$ & 79.5/76 & 3 \\
		Mrk 732 & 0.029 & 7.23$\pm$0.10 & 1.85$^{+0.07}_{-0.07}$ & $\geq$  173 & 7.95$^{+0.31}_{-0.31}$& 269.6/258 & 1 \\
		Mrk 739E & 0.030 & 7.14$\pm$0.10 & 1.87$^{+0.07}_{-0.08}$ & $\geq$  143 & 4.80$^{+0.30}_{-0.29}$ & 113.1/106 & 1 \\
		Mrk 813 & 0.110 & 8.87$\pm$0.10 & 1.85$^{+0.03}_{-0.10}$ & $\geq$  177 & 7.95$^{+0.31}_{-0.31}$ & 269.6/230 & 1 \\
		Mrk 817 & 0.031 & 7.59$\pm$0.07 & 1.65$^{+0.04}_{-0.05}$ & $\geq$  230 & 7.82$^{+0.36}_{-0.30}$ & 263.1/214 & 1 \\
		Mrk 841 & 0.036 & 7.81$\pm$0.10 & 1.78$^{+0.05}_{-0.06}$ & $\geq$  179 & 17.9$^{+1.0}_{-0.7}$ & 403.9/425 & 1 \\
		Mrk 1018 & 0.042 & 8.03$\pm$0.10 & 1.81$^{+0.14}_{-0.35}$ & $\geq$  212 & 1.76$^{+0.46}_{-0.24}$ & 50.5/51 & 1 \\
		Mrk 1044 & 0.016 & 6.44$\pm$0.10  & 1.93$^{+0.05}_{-0.05}$ & $\geq$  214 & 8.74$^{+0.36}_{-0.36}$ & 215.2/205 & 1 \\
		Mrk 1310 & 0.019 & 6.21$\pm$0.08 & 1.77$^{+0.05}_{-0.10}$ & $\geq$  130 & 8.68$^{+0.72}_{-0.32}$ & 215/217 & 1 \\
		Mrk 1393 & 0.054 & 7.87$\pm$0.10  & 1.25$^{+0.23}_{-0.28}$ & $\geq$  19 & 2.07$^{+0.47}_{-0.21}$ & 109.8/79 & 1\\
		NGC 0985 & 0.043 & 7.92$\pm$0.10 & 1.69$^{+0.10}_{-0.11}$ & $\geq$  121 & 11.7$^{+1.1}_{-0.97}$ & 187/195 & 1 \\
		PG 0804+761 & 0.100 & 8.73$\pm$0.05  & 1.69$^{+0.07}_{-0.05}$ & $\geq$  183 & 6.85$^{+0.34}_{-0.33}$ & 126.3/155 & 1 \\
		PKS 0558-504 & 0.137 & 7.33$\pm$0.10 & 2.13$^{+0.05}_{-0.06}$ & $\geq$  134 & 10.4$^{+3.7}_{-0.39}$ & 217.9/206 & 1 \\
		RBS 0295* & 0.074 & 7.97$\pm$0.52 & 1.78$^{+0.10}_{-0.17}$ & $\geq$  49 & 5.32$^{+0.51}_{-0.42}$ & 149.3/153 & 1 \\ 
		RBS 0770 & 0.032 & 7.34$\pm$0.10 & 1.80$^{+0.03}_{-0.03}$ & $\geq$  267 & 22.8$^{+0.59}_{-0.58}$& 400.3/434 & 1 \\
		RBS 1037* & 0.084 & 7.97$\pm$0.52 & 1.88$^{+0.01}_{-0.15}$ & $\geq$  92 & 3.77$^{+0.17}_{-0.19}$ & 180.6/185 & 3 \\
		RBS 1125 & 0.063 & 7.76$\pm$0.20 & 1.86$^{+0.08}_{-0.10}$ & $\geq$  98 & 4.11$^{+0.25}_{-0.25}$ & 109.1/107 & 1 \\
		SBS 1136+594 & 0.060 & 7.98$\pm$0.10 & 1.71$^{+0.06}_{-0.08}$ & $\geq$  92 & 10.6$^{+0.4}_{-0.4}$ & 295/285 & 1 \\
		SDSS J104326.47+110524.2 & 0.048 & 7.91$\pm$0.10 & 1.52$^{+0.24}_{-0.18}$ & $\geq$  34 & 4.13$^{+0.69}_{-0.23}$ & 122/122 & 1 \\
		UGC 06728 & 0.006 & 5.66$\pm$0.10 & 1.57$^{+0.07}_{-0.11}$ & $\geq$  67 & 6.96$^{+0.30}_{-0.30}$ & 227.3/208 & 1\\
		UM 614 & 0.033 & 7.09$\pm$0.10 & 1.64$^{+0.09}_{-0.11}$ & $\geq$  106 & 7.31$^{+0.71}_{-0.50}$ & 162.7/172 & 1\\
		WKK 1263 & 0.024 & 8.25$\pm$0.10 & 1.73$^{+0.04}_{-0.04}$ & $\geq$  224 & 31.6$^{+1.5}_{-1.1}$ & 470.3/503 & 1 \\
		\\ \hline	
		
	\end{tabular*}
\tablecomments{Sources marked with an asterisk (*) correspond to AGN whose black hole masses were taken to be the median black hole mass of the type 1 AGN in the BAT AGN Spectroscopic Survey (BASS) \citep{koss-2017}. \\ $^{A}$ Reference: \citet{koss-2017}. \\ $^{B}$ Unabsorbed 0.1 - 200 keV flux extrapolated from applied spectral model \\ $^{C}$ Applied XSPEC models: (1) \texttt{constant $\times$ TBabs $\times$ zwabs $\times$ cutoffpl} \\ (2) \texttt{constant $\times$ TBabs $\times$ zwabs $\times$ (cutoffpl + pexrav)} \\ (3) \texttt{constant $\times$ TBabs $\times$ zwabs $\times$ (cutoffpl + zgauss)}}	
\end{table*}

\end{document}